\documentclass[preprint,aps,11pt,floatfix,showpacs]{revtex4-1}%
\usepackage{amsmath}
\usepackage{graphicx,epsfig}
\usepackage{amsfonts}
\usepackage{amssymb}
\usepackage[usenames,dvipsnames]{pstricks}
\usepackage{graphicx}%
\providecommand{\U}[1]{\protect\rule{.1in}{.1in}}
\providecommand{\U}[1]{\protect\rule{.1in}{.1in}}
\providecommand{\U}[1]{\protect\rule{.1in}{.1in}}

\begin{document}
\title{Schr\"{o}dinger Equation for Non-Pure Dipole Potential in 2D Systems}
\author{{M. Moumni}$^{1,2}$ and {M. Falek}$^{1,2}$}
\email{m.moumni@univ-biskra.dz}
\affiliation{$^{1}$D\'{e}partement des Sciences de la Mati\`{e}re, Facult\'{e} des Sciences
Exactes \& S.N.V., Universit\'{e} de Biskra, (07000) Biskra, Alg\'{e}rie.}
\affiliation{$^{2}$Laboratoire (L.P.P.N.M.M), Facult\'{e} des Sciences Exactes \& S.N.V.,
Universit\'{e} de Biskra, (07000) Biskra, Alg\'{e}rie.}

\begin{abstract}
In this work, we analytically study the Schr\"{o}dinger equation for the (non-pure) dipolar ion potential $V(r)=q/r+D\cos\theta/r^{2}$, in the case of 2D systems using the separation of variables and the Mathieu equations for the angular part. We give the expressions of eigenenergies and eigenfunctions and study their dependence on the dipole moment $D$. Imposing the condition of reality on the energies $E_{n,m}$ implies that the dipole moment must not exceed a maximum value otherwise the corresponding bound state disappears. We also find that the $s$ states ($m=0$) can no longer exist in the system as soon as the dipole term is present.
\end{abstract}


\maketitle

\section{Introduction}

The solution of the Schr\"{o}dinger equation is very important for the study of atoms and molecules. However the number of systems for which analytical solutions exist is very limited. This motivated the search for simplified models of these physical systems resulting in several studies on spherical potentials. Despite such simplifications only few potentials have been analytically solved. Besides, real physical systems, such as atoms and molecules, rarely exhibit simple spherical symmetry like that of the hydrogen atom.

The study of non-central potentials began with the pioneering works of Makarov \cite{Makarov} and Hartman \cite{Hartmann}, and was then properly structured with the work of Hautot \cite{Hautot}. Their results paved the way for realistic applications of non-central-potential problems such as ring-shaped organic molecules which include, for instance, cyclic polyenes and benzene \cite{Gharbi}. Since then there has been a significant interest in the literature in studies of non-central and ring-shaped potentials (see for instance Refs. \cite{Bharali, Gribakin,Sun} and references therein). However, as demonstrated by Hautot \cite{Hautot}, only few of these potentials in fact have analytical solutions, and they have thus been dominantly considered with either numerical technics or approximation methods.

Owing to the emergence of graphene \cite{Geim,Castro}, two-dimensional systems have recently attracted a lot of interest in the domain of Material Sciences. For this reason, this paper is devoted to the analytical treatment of the dipole potential added to the Coulomb potential $V(r)=a/r+b\cos\theta/r^{2}$ for 2D systems.  This potential, which we call non-pure dipole, is applicable to the case of an ionised dipolar molecule such as water. It has been considered in the case of three-dimensional systems by AlHaidari using the approximation method of tri-diagonal matrices \cite{AlHaidari} and by Moumni et al using perturbation method \cite{Moumni}. Furthermore,
the pure dipole potential $b\cos\theta/r^{2}$ in 3D has been widely studied both in nuclear and molecular physics \cite{Fermi,LeBlond} (see also references in \cite{AlHaidari}), where it has been shown that the dipole moment must exceed a critical value in order to have bound states. On the other hand, in the case of molecular dipole or pure dipole potentials, the moment must be below a critical value in order for bound states to exist \cite{AlHaidari}. For 2D systems, the pure dipole was studied in Refs. \cite{De Martino,Cuenin,Connolly} and it was found in Ref \cite{Connolly} that the critical dipole moment for bound states to exist is zero.

This paper is organised as follows. In section \ref{sec:np}, we start by showing that this potential is the first non-central approximation when considering a non-zero charge distribution which is not spherically symmetric or when considering non-spherical ions. This potential is applicable for most physical systems excluding neutral molecules. We then solve the Schr\"{o}dinger equation for this potential,  in section \ref{sec:SR}, using the separation of variables. Finally we conclude by discussing our results in section \ref{sec:conc}.

\section{Non-pure dipole potential}\label{sec:np}

Our aim is to solve the Schr\"{o}dinger equation for a system consisting of a point charge $q$ under the effect of an extended charge $Q=\sum_j q_{j}$ (a cluster of point charges $q_{j}$). The latter is characterized by a non-zero total charge and a non-spherically-symmetric distribution. One can take as an example of this system a polar ion and a point charge. The potential produced by the charge distribution at the position of the test charge $q$ is written as follows
\begin{equation}
V(r)=\sum_j \frac{1}{4\pi\epsilon_{0}}\,\frac{q_{j}}{r_{j}}\,. \label{1}
\end{equation}
Here $q_{j}$ is the charge of the $j^{\text{th}}$ component of $Q$, and $\vec{r}_j=\overrightarrow{A_jM}=\overrightarrow{OM} -\overrightarrow{OA_j}=\vec{r}-\vec{a}_{j}$ is the position of the point charge $q$ relative to this component. We defined $M$ as the position of the charge $q$ (also denoted by the vector $\vec{r}\,$) and $A_{j}$  as that of $q_j$ (defined by the vector $\vec{a}_j$) relative to the origin $O$, which we choose to coincide with the center of the charge $Q$. Thus we write
\begin{align}
V(r)&=\sum_j\frac{1}{4\pi\epsilon_{0}}\,\frac{q_j}{\left| \vec{r}-\vec{a}_{j}\right|}=\sum_{j}\frac{1}{4\pi\epsilon_0}\, q_j\left[\left(\vec{r}-\vec{a}_j\right)^2\right]^{-1/2}\nonumber\\
& =\sum_j\frac{1}{4\pi\epsilon_{0}}\,q_j\left(\vec{r}\,^2-2\,\vec{r}\cdot\vec{a}_j+\vec{a}_j^{\,2}\right)^{-1/2}\nonumber\\
& =\sum_j\frac{1}{4\pi\epsilon_{0}}\,\frac{q_j}{r}\left(1-2\,\frac{\vec{r}\cdot\vec{a}_j}{\vec{r}\,^2} +\frac{\vec{a}_j^{\,2}}{\vec{r}\,^2}\right)^{-1/2} \,.\label{3}
\end{align}

We assume that the dimensions of the extended charge $Q$ are small compared to those of the whole system constituted by $Q$ and the point charge $q$, such that we write $\left|\vec{a}_j\right|\ll\left|\vec{r}\, \right|$, and thus we have
\begin{equation}
\left(1-2\,\frac{\vec{r}\cdot\vec{a}_j}{\vec{r}\,^2} +\frac{\vec{a}_j^{\,2}}{\vec{r}\,^2}\right)^{-1/2} \simeq 1+\frac{\vec{r}\cdot\vec{a}_{j}}{\vec{r}\,^2}+\mathcal{O}\left(\frac{\vec{a}_j^{\,2}}{\vec{r}\,^2}\right).  \label{4}
\end{equation}
Taking into account these considerations and keeping only the terms up to order $a_j/r$ in the above expansion, one can easily write the potential as a multipolar expansion
\begin{equation}
V(r)=\frac{1}{4\pi\epsilon_{0}}\left(\sum_j\frac{q_j}{r}+\sum_j \frac{q_j a_{j}\cos\theta_j}{r^2}\right). \label{5}
\end{equation}
The first term in this expression is the Coulomb interaction between the total charge $Q$ and the point charge $q$, or the monopole part, while the second is the effect of the geometry of the non-spherically-symmetric charge distribution $Q$, which represents a dipole part. We can simplify the dipole term by mapping the charge distribution $Q$ into a single dipole whose poles are the two centers of all positive and negative charges contained within it. This leads to a dipole-ion potential, i.e. a non-pure dipole whose total charge is non-zero
\begin{equation}
V(r)=\frac{1}{4\pi\epsilon_{0}}\left(\frac{Q}{r}+\frac{Qd\cos\theta}{r^2}\right)=\frac{1}{4\pi\epsilon_{0}}\,\frac{Q}{r}+\frac{1}{4\pi\epsilon_{0}}\,\frac{D\cos\theta}{r^2}\,, \label{6}
\end{equation}
where $d$ is a characteristic distance of the dipole, $\theta$ is the angle between the position vector of $q$ relative to the center of the dipole and the axis of this dipole, and $D=Qd$ is the dipole moment.

\section{2D Schr\"{o}dinger equation for a non-pure dipole potential}\label{sec:SR}

The stationary Schr\"{o}dinger equation reads
\begin{equation}
\left[-\frac{\hbar^2}{2m}\Delta+V\right]  \psi=E\psi\,.\label{7}
\end{equation}
Since we are working with a 2D system and referring to the shape of our potential, we use the polar coordinates $0\leq r<\infty$ and $0\leq\theta\leq2\pi$, and thus we write the wave equation as follows
\begin{equation}
\left[-\frac{\hbar^2}{2m}\left(\frac{\partial^2}{\partial r^2} + \frac{1}{r}\frac{\partial}{\partial r}+\frac{1}{r^2}\frac{\partial^2}{\partial\theta^2}\right)+\frac{q}{4\pi\epsilon_0}\left( \frac{Q}{r}+ \frac{D\cos\theta} {r^2} \right)\right]\psi=E\psi\,.\label{8}
\end{equation}
This expression may then easily be written in the separate form
\begin{equation}
\left[\left(\frac{\partial^2}{\partial r^2}+\frac{1}{r}\frac{\partial}{\partial r}-\frac{2mqQ}{4\pi\epsilon_0\hbar^2}\frac{1}{r}\right) + \frac{1}{r^2}\left(\frac{\partial^2}{\partial\theta^2}-\frac{2mqD} {4\pi\epsilon_0 \hbar^2} \cos\theta\right) \right]  \psi=-\frac{2mE}{\hbar^{2}}\psi\,.\label{9}
\end{equation}
We use the separation of radial and polar variables to write the solution as $\psi(r,\theta)=r^{-1/2}R(r)\Theta(\theta)$, and to split the equation into two parts, angular and radial ones
\begin{subequations}
\begin{align}
\left(\frac{\partial^2}{\partial\theta^2}-\frac{2mqD}{4\pi\epsilon_0\hbar^2}\cos\theta\right) \Theta(\theta) & =E_\theta\Theta(\theta)\,,\label{10}\\
\left[\frac{\partial^2}{\partial r^2}+\left(E_\theta+\frac{1}{4}\right)  \frac{1}{r^2}-\frac{2mqQ}{4\pi\epsilon_0\hbar^2}\frac{1}{r}\right]  R(r)&=-\frac{2mE}{\hbar^{2}}R(r)\,.     \label{11}
\end{align}
\end{subequations}
In this work, we use the same considerations as those for molecular systems: a positive extended charge and a negative point charge, which are equal in magnitude ($q=Q$). Hence
\begin{subequations}\label{eq:12and13}
\begin{align}
\left(\frac{\partial^2}{\partial\theta^2}-E_\theta+\sqrt{2}D\cos \theta\right) \Theta(\theta)&=0\,, \label{12}\\
\left[\frac{\partial^2}{\partial r^2}+\left(E_\theta+\frac{1}{4}\right) \frac{1}{r^2}+\frac{2}{r}+E\right]  R(r)&=0 \,.\label{13}
\end{align}
\end{subequations}
For simplicity we choose $q=-\left|e\right|$, where $e$ is the electric charge (i.e., the point charge is an electron) and, to reduce the length of the expressions, we employ the atomic Rydberg system of units where $2m_e= \hbar = e^2/2=4\pi\epsilon_0=1$.

In order to extract the energy eigenvalues of the system as well as the eigenfunctions $\psi(r,\theta)$, we first solve the angular equation to find the eigenvalues $E_{\theta}$, which may then be used to solve the radial part.

\subsection{Solution of angular equation}

The angular equation can easily be cast in the Mathieu equation form \cite{Mathieu}, by making the substitutions $\theta=2z$, $a=-4E_{\theta}$ and $p=-2\sqrt{2}D$
\begin{equation}
\frac{\partial^2\Theta(z)}{\partial z^2}+\left(a-2p\cos2z\right) \Theta(z)=0\,. \label{14}
\end{equation}
Notice that since the period of $\theta$ is $2\pi$ then that of $z$ is $\pi$, and hence the solutions  to Eq. \eqref{14} are just the cosine-elliptic $ce_{2m}$ and the sine-elliptic $se_{2m+2}$ solutions of the Mathieu equation, where $m$ is a natural number \cite{Abram}. According to Floquet's theorem \cite{Floquet} (or Bloch's theorem \cite{Bloch}), for a given value of the dimensionless parameter $p$, the solutions are periodic only for specific values of the parameter $a$, which are known as the characteristic values. We denote these values corresponding to cosine solutions by $a$ and those related to sine functions by $b$.

Looking at equations \eqref{8} to \eqref{eq:12and13} one can see that, when the dipole effects disappear, i.e., $D\rightarrow0$, our system has as a limiting behaviour the Coulomb system. We therefore require that our solutions must tend to the Coulomb solutions in this limit. This means that we must only keep the cosine solutions and discard the sine ones sine they are not valid solutions when $D\to 0$ for $m=0$, where $m$ is the orbital quantum number, as we shall show later.

\begin{figure}[ht]
\begin{center}
\includegraphics[width=0.67\textwidth]{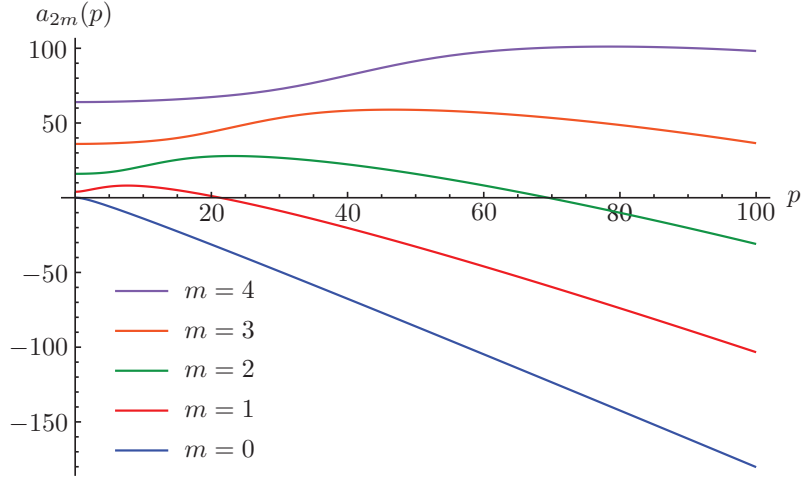}
\caption{\label{Fig1}$a_{2m}(p)$ for $m=0,1,2,3,4$ from bottom to top (in a.u.).}
\end{center}
\end{figure}

In general there is no analytical expression for these solutions and they are usually given either numerically or graphically (see Fig. \ref{Fig1}).
However from these values one can extract the eigen-solutions of the angular equation which can then be used to obtain the radial ones.

Referring to the expression of the dipole potential \eqref{6}, we see that the dipole term is just a correction to the mono-polar term since it is just the latter multiplied by the factor $a/r$. Furthermore, since it depends on the dipole moment, then $p$ can be treated as a small parameter and one can write $p\ll 1$. In this limit, the characteristic values have asymptotic analytical expressions which we write as a series in $p$ \cite{Abram}
\begin{subequations}
\begin{align}
a_0&=-\frac{1}{2}\,p^2+\frac{7}{128}\, p^4-\frac{29}{2304}\, p^6+\mathcal{O}\left(p^8\right), \label{15a}\\
a_2&=4+\frac{5}{12}\,p^2-\frac{763}{13824}\,p^4+\frac{1002401}{79626240}\,p^6+\mathcal{O}\left(p^8\right),\label{15b}\\
a_4&=16+\frac{1}{30}\,p^2+\frac{433}{864000}\,p^4-\frac{5701}{2721600000}\,p^6+\mathcal{O}\left(p^8\right),\label{15c}\\
a_6&=36+\frac{1}{70}\,p^2+\frac{187}{43904000}\,p^4+\frac{6743617}{92935987200000}\,p^6+\mathcal{O}\left(p^8\right).\label{15d}
\end{align}
\end{subequations}
For values of the index of $a_{2m}$ greater than $6$ (i.e., for $m>3$), one can use the same expression for both the $a$'s and the $b$'s, which include up to $\mathcal{O}(p^6)$ terms
\begin{multline}
a_{2m}=4m^2+\frac{1}{2\left(4m^2-1\right)}\,p^2+\frac{20m^2+7}{32\left(4m^2-1\right)^3\left(4m^2-4\right)}\,p^4+\\ +\frac{36m^4+232m^2+29}{64\left(4m^2-1\right)^5\left(4m^2-4\right)\left(4m^2-9\right)}\,p^6+\mathcal{O}\left(p^8\right).\label{16}
\end{multline}
Using the expressions of the characteristic values and the relations $a=-4E_{\theta}$ and $p=-2\sqrt{2}D$, we get the eigenvalues $E_{\theta}$ as a function of the electric moment of the system $D$
\begin{subequations}
\begin{align}
E_\theta^{(0)}&=D^2-\frac{7}{8}\,D^4+\frac{29}{18}\,D^6+\mathcal{O}\left(D^8\right), \label{17a}\\
E_\theta^{(2)}&=-1-\frac{5}{6}\,D^2+\frac{763}{864}\,D^4-\frac{1002401}{622080}\,D^6+\mathcal{O}\left(D^8\right),\label{17b}\\
E_\theta^{(4)}&=-4-\frac{1}{15}\,D^2+\frac{433}{54000}\,D^4+\frac{5701}{21262500}\,D^6+\mathcal{O}\left(D^8\right),\label{17c}\\
E_\theta^{(6)}&=-9-\frac{1}{35}\,D^2-\frac{187}{2744000}\,D^4-\frac{6743617}{726062400000}\,D^6+\mathcal{O}\left(D^8\right).\label{17d}
\end{align}
\end{subequations}
For $m>3$ we have
\begin{multline}
E_\theta^{(2m)}=-m^2-\frac{1}{\left(4m^2-1\right)}\,D^2-\frac{20 m^2+7}{2\left(4m^2-1\right)^3\left(4m^2-4\right)}\,D^4+\\
-\frac{32\left(36m^4+232m^2+29\right)}{\left(4m^2-1\right)^5\left(4m^2-4\right)\left(4m^2-9\right)}\,D^6+\mathcal{O}\left(D^8\right).\label{18}
\end{multline}
We see that in the limit $p\to 0$ (or $D\to 0$), the Mathieu equation has as solutions $\cos m\theta/2$, with $a=4m^2$. Thus the characteristic values can in all cases be written as
\begin{equation}
a_{2m}=4m^2+P_m(p)\,.\label{19}
\end{equation}
Similarly for the angular eigenvalues we have
\begin{equation}
E_\theta^{(2m)}=-m^2+P_m(D)\,,\label{20}
\end{equation}
where $P_m(p)$ and $P_m(D)$ are polynomials which are written in terms of even powers of $p$ and $D$ starting from $2$.

Next we use the expression of $E_\theta^{(2m)}$ in order to solve the radial equation.

\subsection{Solution of the radial equation}

We rewrite the radial equation \eqref{13} in terms of $E_{\theta}^{(2m)}$ as
\begin{equation}
\left[\frac{\partial^2}{\partial r^2}+E+\frac{2}{r}+\left(E_\theta^{(2m)}+\frac{1}{4}\right)\frac{1}{r^2}\right]R(r)=0\,.\label{21}
\end{equation}
In order to simplify Eq. \eqref{21} into a class of known differential equations we use the following ansatz:
\begin{equation}
R\left(r\right)=r^\lambda e^{-\beta r} f\left(r\right)\,,\label{22}
\end{equation}
where $\lambda$ is a constant to be determined. By means of the substitution \eqref{22}, the differential equation for $f\left(r\right)$  becomes
\begin{equation}
\left[  r\dfrac{\mathrm{d}^2\phantom{r}}{\mathrm{d}r^2}+2\left(\lambda-\beta r\right)\dfrac{\mathrm{d}\phantom{q}}{\mathrm{d}q}+2\left(1-\lambda\beta\right)\right]f=0\,,\label{23}
\end{equation}
where $\beta$ and $\lambda$ satisfy the following relations
\begin{equation}
\beta^2=-E\,,\qquad \text{and}\qquad \lambda\left(\lambda-1\right)+E_{\theta}^{(2m)}+\frac{1}{4}=0 \,.\label{24}
\end{equation}
Solving for $\lambda$ in the latter equation yields two solutions
\begin{equation}
\lambda=\frac{1}{2}\pm\sqrt{-E_\theta^{(2m)}}\,.\label{25}
\end{equation}
However since we require $R\left(r\right)$ to be a nonsingular function at $r=0$, then the accepted value of $\lambda$ is
\begin{equation}
\lambda=\frac{1}{2}+\sqrt{-E_\theta^{(2m)}} \,.\label{26}
\end{equation}

Now taking $z=2\beta r$ and substituting into Eq. \eqref{23}, the latter reduces to a differential equation of the confluent hypergeometric type
\begin{equation}
\left[z \dfrac{\mathrm{d}^2\phantom{z}}{\mathrm{d}z^2}+\left(2\lambda-z\right)\dfrac{\mathrm{d}\phantom{z}}{\mathrm{d}z}-\left(\lambda-\dfrac{1}{\beta}\right)\right]f\left(z\right)=0\,.\label{27}
\end{equation}
The solution of this differential equation which is regular at the origin $z=0$ is given in terms of confluent hypergeometric functions as
\begin{equation}
f(z)=N_1 F_1\left(\lambda-\beta^{-1},2\lambda,z\right),\label{28}
\end{equation}
with $N$ a normalization constant to be determined later.

In terms of the variables $r$ and $\theta$, we can now write the general form of the wave function $\psi$ as follows
\begin{equation}
\psi\left(r,\theta\right)=N r^{\lambda-\frac{1}{2}}\,e^{-\beta r}\,\Theta\left(\theta\right)\, _1F_1\left(\lambda-\beta^{-1},2\lambda,2\beta r\right).\label{29}
\end{equation}
The functions $_1F_{1}\left(-n_{r},2\lambda,2\beta r\right)$ may be written as Laguerre polynomials of degree $n_r$ as follows
\begin{equation}
L_{n_r}^{\left(2\lambda-1\right)}\left(2\beta r\right)=\frac{\left(n_r+2\lambda-1\right)!}{n_r!\left(2\lambda-1\right)!}\,_1F_1\left(-n_r,2\lambda,2\beta r\right).\label{30}
\end{equation}

To determine the normalization constant $N$ we substitute the wavefunction \eqref{29} into the normalization condition $\int \left\vert\psi\left(r,\theta\right)\right\vert^2 r\mathrm{d}r\,\mathrm{d}\theta=1$,
where we recall that $\Theta\left(\theta\right)$ is the Mathieu solution $ce_{2m}(\theta/2)$ which is normalized by definition \cite{Abram}. Using the identity \cite{Grad}
\begin{equation}
\int_0^\infty e^{-q}\,q^{k+1} \left[L_n^k\left(q\right)\right]^2 \mathrm{d}q=\frac{\left(n+k\right)!}{n!}\left(2+k+1\right),\label{31}
\end{equation}
we obtain
\begin{equation}
N=\frac{2^\lambda\beta^{\lambda+\frac{1}{2}}}{\left(2\lambda-1\right)!}\left[\frac{\left(n+2\lambda-1\right)!}{n!\left(n+\lambda\right)}\right]^{\frac{1}{2}}.\label{32}
\end{equation}

Using the condition of the convergence of the solutions at infinity, and from the asymptotic behavior of the confluent series $_1F_1$ of Eq. \eqref{29}, that for $r\to\infty$ we have $_1F_1\to 0$ which leads to $\psi\to 0$ at infinity, we obtain the following general quantum condition
\begin{equation}
\lambda-\beta^{-1}=-n_r\,,\qquad n_r=0,1,2,\hdots \,,\label{33}
\end{equation}
and we get the discrete energy levels from the condition \eqref{33} as follows
\begin{equation}
E_{n_r,m}=-\left(n_r+\sqrt{-E_\theta^{(2m)}}+\frac{1}{2}\right)^{-2}\,.\label{34}
\end{equation}
We can relate these energies with the Coulomb energy by using the relationship \eqref{20}
\begin{equation}
E_{n_r,m}=-\left(n_r+\sqrt{m^2-P_m(D)}+\frac{1}{2}\right)^{-2}\,,\label{35}
\end{equation}
and the limit $P_m(D)\to 0$ gives us
\begin{equation}
E_{n_r,m}=-\left(n_r+\left\vert m\right\vert-\frac{P_m(D)}{2}+\frac{1}{2}\right)^{-2}=-\left(n+\frac{1}{2}-\frac{P_m(D)}{2}\right)^{-2}\,,\label{36}
\end{equation}
which are the energies for the Coulomb potential with $n=n_r+\left\vert m\right\vert$, where we have the condition $m\leq n$ from the Coulomb system \cite{Zaslow,Parfitt}. The final expression for the
energy eigenvalues is
\begin{equation}
E_{n,m}=-\left(n-\left\vert m\right\vert+\sqrt{-E_\theta^{(2m)}}+\frac{1}{2}\right)^{-2}\,.\label{37}
\end{equation}

The requirement that these energy eigenvalues $E_{n,m}$ be real necessitates that $E_\theta^{(2m)}$ be negative, and thus from the relation $a=-4E_\theta$ the characteristic values must be positive. This gives us a condition on the parameter $p$ and therefore on the dipole moment $D$ from the definition $p=-2\sqrt{2}D$. Consequently, we conclude that in order for bound states $E_{n,m}$ to exist, it is necessary that the dipole moment $D$ not exceed a critical value determined by the equation $E_\theta^{(2m)}=0$. This critical value depends only on $m$ and is thus denoted $D_{\mathrm{crit}}^{(m)}$. We show in table \ref{table} the values $D_{\mathrm{crit}}^{(m)}$ (in Rydberg atomic units) for some values of $m$.
\begin{table}[ht]
\centering
\begin{tabular}[c]{|c|c|c|c|c|c|c|c|c|}\hline
$m$ & $0$ & $1$ & $2$ & $3$ & $4$ & $5$ & $6$ & $7$\\\hline
$D_{\mathrm{crit}}^{(m)}$ & $0,000$ & $7,530$ & $24,547$ & $51,285$ & $87,746$ & $133,930$ & $189,837$ & $255,468$\\\hline
\end{tabular}
\caption{\label{table}Critical value $D_{\mathrm{crit}}^{(m)}$  for some values of $m$.}
\end{table}
We note that when the dipole moment is non-zero then the level $E_{0,0}$ disappears. We shall elaborate on this point separately in the next section.

\begin{figure}[ht]
\begin{center}
\includegraphics[width=0.53\textwidth]{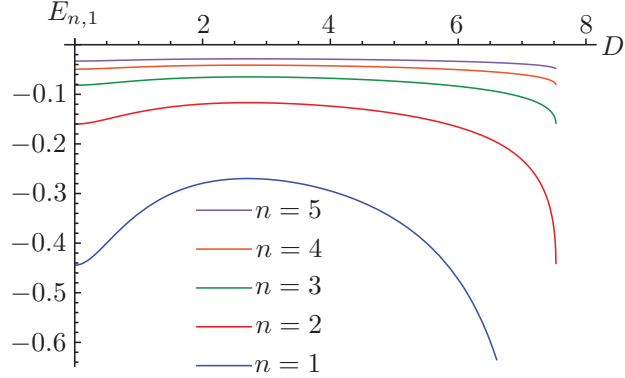}
\caption{\label{Fig2}$E_{n,1}(D)$ for $n=1,2,3,4,5$ from bottom to top (in a.u.).}
\end{center}
\end{figure}

\begin{figure}[ht]
\begin{center}
\includegraphics[width=0.53\textwidth]{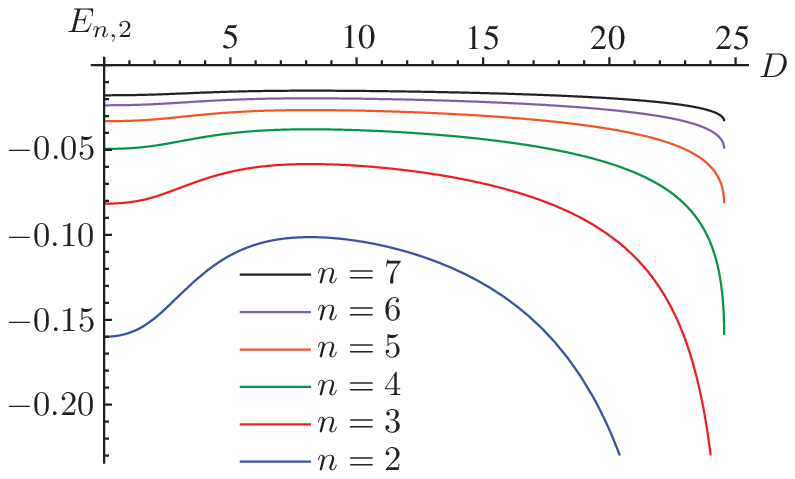}
\caption{\label{Fig3}$E_{n,2}(D)$ for $n=2,3,4,5,6,7$ from bottom to top (in a.u.).}
\end{center}
\end{figure}

\begin{figure}[ht]
\begin{center}
\includegraphics[width=0.8\textwidth]{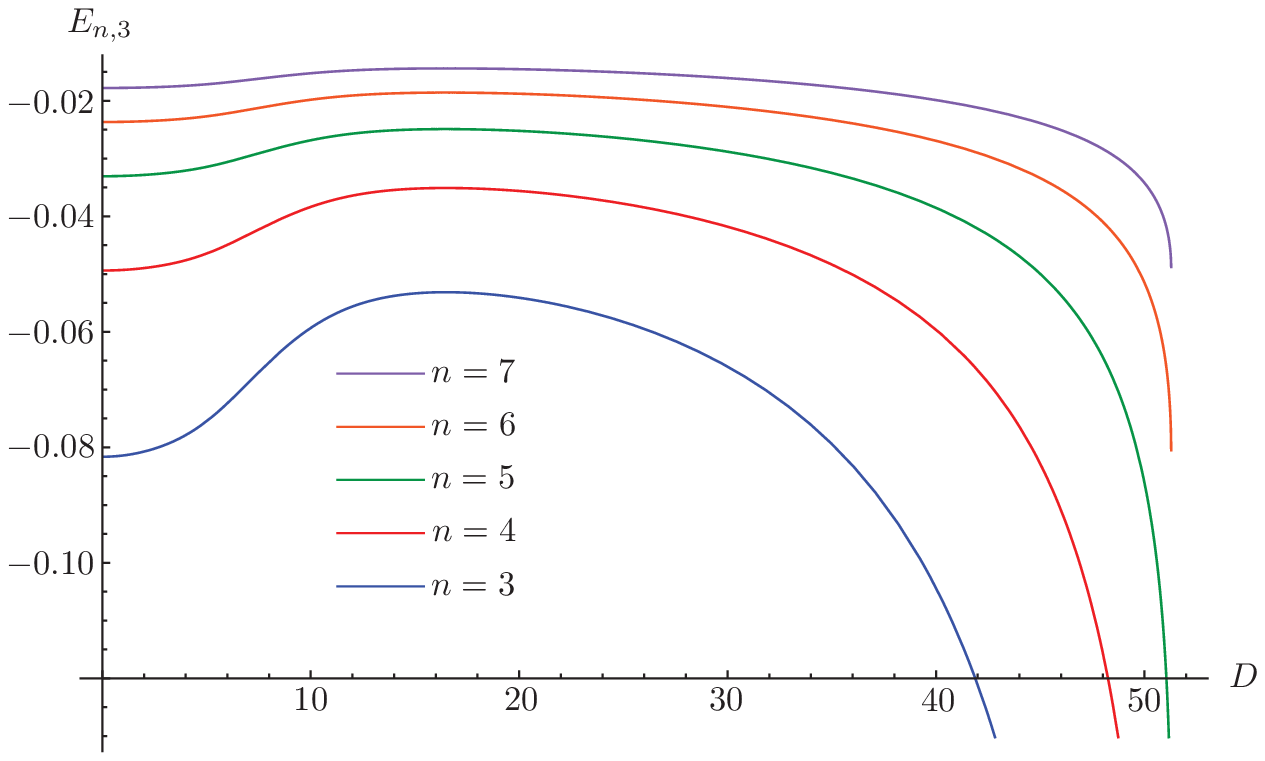}
\caption{\label{Fig4}$E_{n,3}(D)$ for $n=3,4,5,6,7$ from bottom to top (in a.u.).}
\end{center}
\end{figure}

\begin{figure}[ht]
\begin{center}
\includegraphics[width=0.8\textwidth]{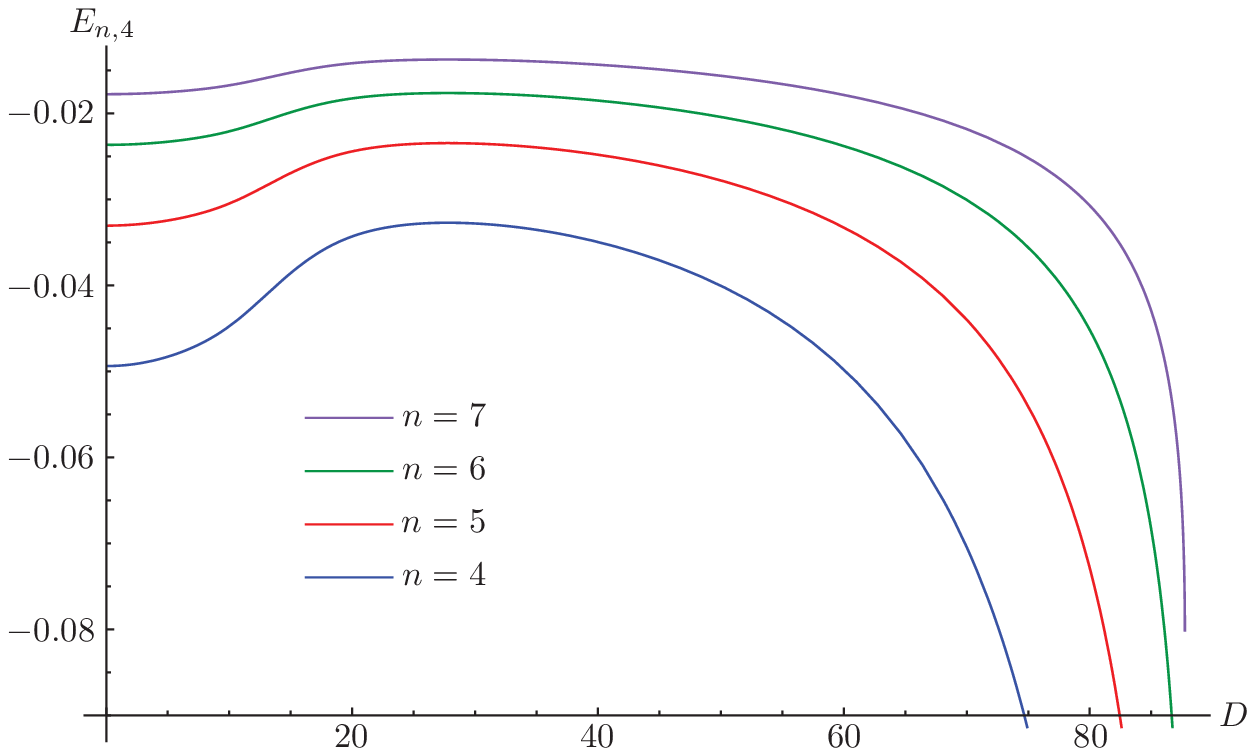}
\caption{\label{Fig5}$E_{n,4}(D)$ for $n=4,5,6,7$ from bottom to top (in a.u.).}
\end{center}
\end{figure}

From the relation \eqref{34}, we can plot all the graphs for the energies for all possible values of the dipole moment $E_{n,m}(D)$, as shown in figures \ref{Fig2}, \ref{Fig3}, \ref{Fig4} and \ref{Fig5}.

\section{Conclusions}\label{sec:conc}

In this work, we have analytically studied the Schr\"{o}dinger equation for the potential $V(r)=a/r+b\cos\theta/r^{2}$  for 2D systems using the separation of variables and the Mathieu equations for the angular part. We
have shown that this potential is applicable to a dipolar ion, and that it is the first-order approximation resulting from the effect of a non-spherical distribution of charges. We gave the expressions of eigenenergies and eigenfunctions and we studied their dependence on to the dipole moment $D$. As expected, our solutions tend to the Coulomb ones when the dipole moment vanishes.

The plots of the energies show that they increase with the dipole moment up to a maximum value and then start decreasing. The behavior of these solutions is similar to that of the characteristic values of the Mathieu functions. Furthermore, the requirement that the energies $E_{n,m}$ be real implies that the dipole moment must not exceed a maximum value, otherwise the corresponding bound states disappear. These critical values depend only on the magnetic quantum number $m$, and are thus labeled $D_{\mathrm{crit}}^{(m)}$. This result is similar to that found in 3D case by AlHaidari \cite{AlHaidari}. It is also in contrast to the case of a pure dipole (i.e., when no Coulomb term is present), where it is necessary that the dipole moment exceeds a minimum value in order for bound states to exist \cite{Fermi}.

We also found that the $s$-states ($m=0$) no longer exist for this system when the dipole term is present since $D_{\mathrm{crit}}^{(0)}=0$. Noting that the $s$-states exist for other non-central potentials (see for instance Refs. \cite{Bharali,Sun} and references therein), we deduce that the absence of these states in our case is not due to the difference between the symmetry of the $s$-states (which are central) and that of the potential (which is not spherically symmetric). This phenomena requires further studies given the specificity of the potential $\cos\theta/r^{2}$ \cite{Camblong,Coon}. From a purely mathematical point of view, this is due to the fact that the characteristic value $a_0$ is negative for all possible values of the parameter $p$, while the other parameters $a_{m\neq0}$ start from positive values and end in the negative domain.

\end{document}